\documentclass[12pt,twoside]{article}
\usepackage[T1]{fontenc}
\usepackage[latin1]{inputenc}
\usepackage{amsmath}
\usepackage{amssymb}
 \normalsize
\newcommand{\Zint}{\mathbb{Z}}
\newcommand{\Real}{\mathbb{R}}
\newcommand{\axion}{\hbox{\Large $a$} }
\def\Im{\,{\rm Im}\, }
\def\Re{\,{\rm Re}\, }
\makeatletter
\renewcommand{\@makefnmark}{\mbox{$^{\ddagger\@thefnmark}$}}
\makeatother
\newcommand{\publititle}[8]
{ 
  \begin{flushright} #1 \\ {\tt #2} \end{flushright}
  \vfill
  \begin{center}{\Large\scshape\bfseries #3}\end{center}
  \vskip 8mm
  \begin{center}{\large #4}\end{center}
  \begin{center}{\normalsize\sl #5}\end{center}
  \vskip 8mm     
  \nopagebreak 
  \noindent #6
  \vfill 
  \begin{flushleft} #7
  \end{flushleft}
  \hrule width 6.7cm \vskip1mm
  {\small #8}
  \thispagestyle{empty}
  \clearpage
}

\begin{document}

\publititle{CPTH-S604.0498}{hep-th/9804023}
{A note on non-perturbative $R^4$ couplings$^\star$}{B. Pioline}
{Centre de Physique Th{\'e}orique, Ecole Polytechnique,$^\dagger$
  {}F-91128 Palaiseau, France }
{
Exact non-perturbative results have been conjectured for
$R^4$ couplings in type II maximally supersymmetric
string theory. Strong evidence
has already been obtained, but contributions of cusp forms,
invisible in perturbation theory, have remained an open
possibility. In this note, we use the $D=8$ $N=2$ superfield formalism
of Berkovits to prove that supersymmetry requires the exact $R^4$ threshold 
to be an eigenmode of the Laplacian on the scalar manifold
with a definite eigenvalue. Supersymmetry and U-duality invariance
then identify the exact result with the order-$3/2$ Eisenstein
series, and rule out cusp form contributions. 
}
{CPTH-S604.0498\\April 1998, to appear in Phys. Lett. B.}
{ $^{\star}$ {\small Research supported in part by the EEC 
under the TMR contract ERBFMRX-CT96-0090.}\\
$^{\dagger}${\small Unit{\'e} mixte du CNRS UMR 7644.} 
}

\clearpage

Following the original conjecture by Green and Gutperle\cite{Green:1997tv},
$R^4$ couplings in maximally supersymmetric 
type II strings have received much attention over 
the last year. The conjecture has been extended in lower
dimensions \cite{Green:1997di,Kiritsis:1997em,Pioline:1997pu}
and for other eight-derivative terms in the effective action
\cite{Kehagias:1997cq,Green:1997me} as well as for an 
infinite class of higher-derivative topological couplings
\cite{Russo:1997mk,Russo:1997fi,Berkovits:1998}. 
With the possible exception of Refs.~\cite{Russo:1997mk,Russo:1997fi},
all these terms share the property
of 1/2-BPS saturation, or in the language of superspace derive from
chiral lagrangian densities. Expanded at weak coupling,
the $R^4$ threshold exhibits in addition to the tree-level and one-loop
terms an infinite sum of non-perturbative effects
that can be interpreted as coming from euclidean D-branes
wrapped on supersymmetric cycles of the compactification
manifold \cite{Kiritsis:1997em,Pioline:1997pu,Pioline:1997ix}.
This gives much insight into the rules of semiclassical calculus
in string theory, and by itself is a strong support in favor of the 
conjecture. Stronger support still has come from
Ref.~\cite{Antoniadis:1997zt}, in which the D=10 type IIB
$R^4$ threshold was derived from heterotic-type II duality in $N=4$ $D=4$
although in a not too direct way; and from Ref.~\cite{Berkovits:1997pj},
in which N=2 D=8 superspace technics were developped and used to
show that the $R^4$ couplings were not corrected {\it perturbatively},
in agreement with the conjectures of 
Refs.~\cite{Green:1997tv,Kiritsis:1997em}.

Despite these arguments, the possibility has remained that the
$Sl(2,\Zint)$ (or $Sl(3,\Zint)$) Eisenstein series $E_s$ with $s=3/2$
may not be the full answer for the exact type IIB ten-dimensional
(or eight dimensional) $R^4$ coupling 
$f_{10}$ (or $f_{8}$), due to the existence of a discrete
series of {\it cusp forms} $E_{s_n}$ invariant under 
the U-duality group, but contributing only at the non-perturbative 
$O(e^{-1/g})$ level; no explicit example of cusp form
(at least for $Sl(2,\Zint)$ or $Sl(3,\Zint)$) is known in
the mathematical literature, and it would be very interesting
to find such examples from string theory. 
Eisenstein series and cusp forms are the only eigenmodes
of the Laplace operator on the fundamental domain
\begin{equation}
\label{lapsl2}
\Delta E_s = s(s-1) E_s\ ,\quad 
\Delta=4\tau_2^2 \partial_\tau\partial_{\bar\tau}
\end{equation}
in the $Sl(2,\Zint)$ case \cite{Terras:1985}, or
\begin{equation}
\label{eigsl3}
\Delta E_s = \frac{2s(2s-3)}{3} E_s\ ,
\end{equation} 
\begin{equation} 
\label{lapsl3}
\Delta=4\tau_2^2 \partial_\tau\partial_{\bar\tau}
+\frac{1}{\nu\tau_2}|\partial_{B_N}-\tau\partial_{B_R}|^2
+3\partial_{\nu}(\nu^2\partial_{\nu})
\end{equation}
in the $Sl(3,\Zint)$ case \cite{Kiritsis:1997em}. In the above
equations, $\tau=\axion+i~e^{-\phi}$ is the ten-dimensional
complex coupling constant, $\nu=1/(\tau_2 V^2)$ encodes the
volume $V$ of the 2-torus in a $Sl(2,\Zint)_{\tau}$ invariant
way, and $(B_R,B_N)$ is the doublet of expectation values
of the Ramond and Neveu-Schwarz two-forms  on $T^2$.
It has been contemplated by Green and Vanhove that equation
(\ref{lapsl2}) for $s=3/2$ may come as a consequence of
$N=2$ $D=10$ supersymmetry \cite{Green:1997di}; this would
then rule out contributions from cusp forms $E_{s_n}$,
since the latter occur only on the axis $s_n\in 1/2 + i\Real$.
In this note, we wish to prove that this is actually the
case, namely that equation (\ref{eigsl3}) with $s=3/2$ follows
from $N=2$ $D=8$ supersymmetry. Equation (\ref{lapsl2}) 
for $s=3/2$ then follows by decompactification of the two-torus,
since after taking proper account of the Weyl rescalings,
$f_{10}(\tau)=\lim_{\nu\rightarrow 0} \nu^{-1/2} f_{8}(\tau,\nu,B_N,B_R)$. 

Berkovits has given a construction of $D=8$ $R^4$ couplings
in a on-shell linearized superfield formalism
\cite{Berkovits:1997pj}~: the (only) gravitational
multiplet can be described by two mutually constrained 
superfields~: a {\it chiral} superfield
$W$ with lowest component $w$ parametrizing
the $Sl(2,\Real)/U(1)$ factor of the moduli space near the origin, and
a {\it linear} superfield $L_{ijkl}$, with the five lowest components
$l_{ijkl}$ parametrizing the $Sl(3,\Real)/SO(3)$ coset factor
near the origin\footnote{We follow the notations of
Ref.~\cite{Berkovits:1997pj}~; the indices $i,j,\dots=\pm$ 
transform as a doublet of the R-symmetry $SU(2)_R$, identified 
with the $SO(3)$ group in the $Sl(3,\Real)/SO(3)$ coset. Small letters
denote the lowest components of the corresponding superfields.}.
Chiral densities can be formed from each of the two multiplets~:
\begin{equation}
\label{chidens}
\int d^8 x\vert_{\theta^{\alpha}_j=\bar\theta^{\bar\alpha}_j=0}
\left[ (D_+)^8 (D_-)^8 f_V(W)
+\int d\zeta (D_-)^8 (\bar D^{+})^8 f_T(\tilde L,\zeta) 
\right]+c.c.
\end{equation} 
where
\begin{equation}
\tilde L = L_{++++} -4\zeta L_{+++-}+6\zeta^2 L_{++--}
-4\zeta ^3 L_{+---} + \zeta ^4 L_{----}\ .
\end{equation}
Each density is integrated on a different half of the
superspace generated by the 32 odd coordinates $\theta_i ^{\alpha},
\bar\theta_i ^{\bar\alpha}$, and generates sixteen-fermion
or eight-derivative 1/2-BPS saturated interactions. 
The first term describes among other terms the $R^4_{++}=\left(t_8 t_8 -
\frac{1}{4}\epsilon_{8}\epsilon_{8}\right)R^4$ coupling,
in terms of a holomorphic fonction $f_V(w)$, while the second 
describes\footnote{It also describes the other terms investigated
in Refs. \cite{Kehagias:1997cq,Green:1997me} which are
related to $R_{+-}^4$ by supersymmetry.} the 
$R^4_{+-}=\left(t_8 t_8+\frac{1}{4}\epsilon_{8}\epsilon_{8}\right)R^4$ coupling
in terms of a  function $f_T(\zeta,\tilde l)$ holomorphic in
the first variable~:
\begin{equation}
\label{sr4}
S_{R^4} = \int d^8 x \sqrt{g} 
\left[ \partial^4_w f_V(w)~ R_{++}^4 
+ \int d\zeta ~\partial^4_{\tilde l} f_T(\zeta,\tilde l) ~R_{+-}^4
\right]
\end{equation}
Invariance under
supersymmetry of the second term in equation (\ref{chidens}) 
is guaranteed by the constraints
\begin{equation}
D_{(i}^\alpha L_{jklm)} = \bar D_{(i}^{\bar \alpha} L_{jklm)} = 0
\end{equation}
which imply
\begin{equation}
(D_+ -\zeta D_{-}) \tilde L= (\bar D^- - \zeta \bar D^+) \tilde L =0
\ .
\end{equation}
The scalar fields $l_{ijkl}$ can be recast as a symmetric traceless
representation of $SO(3)$, real thanks to the constraint
$(L_{ijkl})^* = L^{ijkl}$~:
\begin{equation}
\begin{pmatrix}
\Re l_{++++} + l_{++--} & \Im l_{++++} & 2 \Re l_{+++-} \\
\Im l_{++++}            & -\Re l_{++++} + l_{++--} & -2 \Im l_{+++-} \\
2\Re l_{+++-}           & -2 \Im l_{+++-} & -2 l_{++--} 
\end{pmatrix}\ ,
\end{equation}
which is nothing but the symmetric
determinant 1 $Sl(3,\Real)$ matrix $M$ parametrizing the coset
space $Sl(3,\Real)/SO(3)$ {\it close to the origin}\footnote{This
is not a restriction, since the moduli space is an homogeneous space
anyway.}~:
\begin{align}
M=&\frac{\lambda^{2/3}}{T_2}
\begin{pmatrix}
1   & T_1 & -\Re B \\
T_1 & |T|^2 & \Re(\bar T B)\\
-\Re B& \Re(\bar T B)& \frac{T_2}{\lambda^2} + |B|^2
\end{pmatrix}\\
\sim&
\begin{pmatrix}
1-\hat T_2 +\frac{2}{3}\hat\lambda   & \hat T_1 & -\Re \hat B \\
\hat T_1 & 1 + \hat T_2 + \frac{2}{3} \hat \lambda & \Im\hat B\\
-\Re \hat B& \Im\hat B& 1-\frac{4}{3}\hat\lambda
\end{pmatrix}
\end{align}
where $T= B_N + i~V$, $\lambda = 1/(\tau_2 V^{1/2})$ is the 
T-duality invariant dilaton, $B=-B_R+i~V\axion$\footnote{
This parametrization can be obtained from the one in 
Ref.~\cite{Kiritsis:1997em} by an automorphism
$(\tau,\nu,B_R+\tau B_N)\rightarrow(T,\lambda^2,-B_R+i~V\axion)$. It has
the advantage of putting the dependance on the RR zero-modes
in a single complex field.}, and the hat denotes the variation
away from the origin $M=\mathbb{I}$.
One therefore identifies
\begin{equation}
l_{++++} = i \hat T\,\quad
l_{+++-} = - \hat B/2\ ,\quad
l_{++--} = \frac{2}{3} \hat \lambda\ .
\end{equation}
As noted by Berkovits, the {\it perturbative} contribution to the
$R^4$ couplings has to be independent of the complex RR zero-mode
$\hat B$ (that is under the continuous Peccei-Quinn symmetries
$B \rightarrow B+ {\rm cste}$). 
This restricts the {\it perturbative} function $f_{T}$ to
\begin{equation}
f_{T}^{\rm pert.}(\zeta,\tilde l)= h \frac{\tilde l^5}{\zeta^3}
+\frac{g(\tilde l)}{\zeta}\ , \quad h\in \Real
\end{equation}
which describes the tree-level and one-loop contributions 
respectively \cite{Berkovits:1997pj}. 
Our aim here is however to extend this argument
to the non-perturbative level.
In terms of the variables $(\lambda,T,B)$, 
the Laplacian (\ref{lapsl3}) indeed takes the form
\begin{equation} 
\label{lapsl3b}
\Delta=4T_2^2 \partial_T\partial_{\bar T}
+4 \lambda^2 T_2 \partial_B \partial_{\bar B}
+\frac{3}{4\lambda}\partial_{\lambda}(\lambda^3\partial_{\lambda})
\end{equation}
and linearizes around $M\sim \mathbb{I}$ to
\begin{equation} 
\label{linlap}
\Delta_{\rm lin}=
\frac{3}{4} \partial_{\hat\lambda} ^2 + 
4 \partial_{\hat T}\partial_{\bar{\hat T}}
+4 \partial_{\hat B}\partial_{\bar{\hat B}}\ .
\end{equation}
The $R_{+-}^4$ coupling in Eq.\ (\ref{sr4}) is therefore given by
\begin{equation}
f_8 = \int d\zeta~\partial_{i\hat T}^4 
f_T \left(\zeta,~i\hat T + 2\zeta \hat B +4 \zeta ^2 \hat \lambda
-2\zeta ^3 \hat B -i \zeta ^4 \bar{\hat T} \right)
\end{equation}
which is easily seen to be annihilated by the linearized Laplacian
(\ref{linlap}). This analysis around $M\sim \mathbb{I}$ can be carried
out at any point in the $Sl(3,\Real)/SO(3)$ coset, although the
function $f_T$ is only locally defined, and the linearized second
order differential operator carries over to the full $Sl(3,\Real)$
invariant Laplace operator. 
This implies that
{\it the exact $R_{+-}^4$ threshold in type IIB string theory
compactified on $T^2$ is annihilated by the Laplacian}
(\ref{lapsl3}) or (\ref{lapsl3b}).
Actually, this analysis does not take into account the non-local
counterterms familiar for thresholds in the critical dimension
\cite{Dixon:1991pc},
and one should allow for a non vanishing {\it constant} value of the 
Laplacian. Indeed, the $R^4$ threshold conjectured 
in Ref.\ \cite{Kiritsis:1997em} is a quasi-zero-mode of
the Laplacian~:
\begin{equation}
\Delta \hat E_{3/2} = 4\pi\ ,
\end{equation}
and the second member arises from the regularization of 
a logarithmic divergence in
the order-$3/2$ Eisenstein series.  Under decompactification
of the two-torus, one obtains an eigenvector
of the $Sl(2,\Zint)\backslash Sl(2,\Real)/ U(1)$ Laplacian
with eigenvalue $3/4$, and this unambiguously
identifies the $D=10$ $R^4$
threshold with the order-$3/2$ $Sl(2,\Zint)$ Eisenstein
series. Should there be another $Sl(3,\Zint)$ invariant
quasi-zero-mode of the Laplacian in $D=8$, it would
have to reduce to this $Sl(2,\Zint)$ Eisenstein series
in all $Sl(2,\Real)/U(1)\subset Sl(3,\Zint)/SO(3)$
decompactification limits, and this excludes such an
eventuality. The conjectures
of Refs. \cite{Green:1997tv,Green:1997di,Kiritsis:1997em} 
are therefore proved.
The $N=2$ $D=8$ superfield description
of the $R^4 H^{4g-4}$ topological terms 
is not well developped yet, but it is likely that the present method
could be extended to prove that these couplings are indeed given by the 
$E_{g+\frac{1}{2}}$ Eisenstein series, as conjectured
in Ref. \cite{Berkovits:1998}. This leaves open the hunt
for cusp forms in other string theory couplings.

\providecommand{\href}[2]{#2}\begingroup\raggedright\endgroup


\end{document}